\documentclass[11pt]{article}
\usepackage[a4paper,margin=2cm]{geometry}
\usepackage{amsmath, amssymb, amsfonts}
\usepackage[T1]{fontenc}
\usepackage{color}
\usepackage{bm}
\usepackage{subfigure}
\usepackage{graphicx}
\usepackage{multirow}
\usepackage{setspace}
\usepackage{paralist}
\usepackage{float}
\usepackage{comment}
\usepackage{cite}
\usepackage{hyperref}

\usepackage[small,nohug,heads=LaTeX]{diagrams}
\diagramstyle[labelstyle=\scriptstyle]

\definecolor{blueblack}{rgb}{0,0,.7}

\newcommand{\emphdef}[1]{%
  \textcolor{blueblack}{%
    \textbf{\emph{#1}}%
  }%
}
\makeatletter
\def\proof{\@ifnextchar[
  {\@xproof}{\@proof}}
\def\endproof{\unskip\kern 10\p@ \begingroup \unitlength\p@
  \linethickness{.4\p@} \framebox(6,6){} \endgroup \endtrivlist }
\def\@proof{\trivlist \item[\hskip\labelsep {\em Proof.}  ]
  \ignorespaces }
\def\@xproof[#1]{\trivlist \item[\hskip\labelsep{\em #1.}]
  \ignorespaces }
\makeatother

\usepackage{hyperref}

\long\def\myfigure#1{#1} 

\newcommand{\Rspace}        {\mathbb R}

\newcounter{blah} \numberwithin{blah}{section}
\newtheorem{Theorem}[blah]{Theorem}
\newtheorem{Fact}[blah]{Fact}
\newtheorem{Lemma}[blah]{Lemma}
\newtheorem{Proposition}[blah]{Proposition}
\newcommand{\inv}{^{-1}}

\newenvironment{problem}[1]{\medskip\par\parindent 0pt{\textcolor{blueblack}{\bf #1}}}{\medskip\par} 
\newcommand\Input{\textsc{Input}}
\newcommand\Output{\textsc{Output}}

\newcommand\Question{\textsc{Question}}

\newcommand\BdContract{\textsc{BdContract}}
\newcommand\Unknot{\textsc{Unknot}}

\begin{document}

\title{Deciding contractibility of a non-simple curve on the boundary of a 3-manifold: A computational Loop Theorem\thanks{This work is supported by the French ANR Blanc projects ANR-12-BS02-005 (RDAM), ANR-17-CE40-0033 (SoS), and ANR-19-CE40-0014 (Min-max).  A preliminary version (under slightly different title) appeared in \emph{Proceedings of the ACM--SIAM Symposium on Discrete Algorithms (SODA), 2017, pp. 2691--2704.}}}

\author{\'Eric Colin de Verdi\`ere\thanks{LIGM, CNRS, Univ Gustave Eiffel, ENPC, ESIEE Paris, F-77454 Marne-la-Vall\'ee, France.  Email: \texttt{eric.colindeverdiere}\texttt{@u-pem.fr}.  Part of this work was done while this author was at CNRS, 
D\'epartement d'Informatique, \'Ecole normale sup\'erieure, Paris, France.}%
\and Salman Parsa\thanks{Department of Computer Science and Mathematics, Saint Louis University, Saint Louis, MO, USA. Email: \texttt{sparsa@slu.edu}.  Part of this work was done while this author was at Fondation Sciences Math\'ematiques de Paris and D\'epartement d'Informatique, \'Ecole normale sup\'erieure, Paris, France.}}
\date{}
\maketitle

\begin{abstract}
  We present an algorithm for the following problem. Given a triangulated $3$-manifold $M$ and a (possibly non-simple) closed curve on the boundary of $M$, decide whether this curve is contractible in $M$.  Our algorithm runs in space polynomial in the size of the input, and (thus) in exponential time.  This is the first algorithm that is specifically designed for this problem; it considerably improves upon the existing bounds implicit in the literature for the more general problem of contractibility of closed curves in a $3$-manifold. The proof of the correctness of the algorithm relies on methods of $3$-manifold topology and in particular on those used in the proof of the Loop Theorem.  As a byproduct, we obtain an algorithmic version of the Loop Theorem that runs in polynomial space, and (thus) in exponential time.
\end{abstract}

\section{Introduction}

Decision problems related to the homotopy of curves have been considered from combinatorial and algorithmic viewpoints for more than a century.  On surfaces, Dehn~\cite{d-tkzf-12} describes a combinatorial procedure for the \emph{contractibility problem}: Given a closed curve on a surface, decide whether it is contractible.  Dehn's algorithm can be analyzed~\cite{da-cdawp-85}, and is a fundamental tool in geometric group theory~\cite{g-dacwp-60}.  Much more recently, Lazarus and Rivaud~\cite{lr-hts-12} and Erickson and Whittlesey~\cite{ew-tcsr-13} develop linear-time algorithms for the contractibility problem (and the more general free homotopy problem) on a surface.

Homotopy questions of curves turn out to be much more complicated for more general topological spaces.  Indeed, the problem of contractibility in 2-dimensional simplicial complexes is undecidable; this follows from the facts that every finitely presented group is the fundamental group of a 2-complex, which can be constructed easily from the presentation, and that, given a finitely presented group specified by generators and relations, there is no algorithm to decide whether a word in the generators and their inverses represents the identity in the group.  This implies that the contractibility problem is also undecidable for 4-manifolds.  See, e.g., Stillwell~\cite{s-ctcgt-93} for more details. 

Between surfaces and 4-manifolds lies the case of 3-manifolds. All known algorithms for deciding contractibility of curves in 3-manifolds use the Geometrization Conjecture of Thurston and its proof. The first such algorithm is deduced from the fact that the fundamental group of a 3-manifold is residually finite~\cite{Hem87}.  An algorithm also follows from the bounds on the isoperimetric functions of the fundamental groups of 3-manifolds, which are at most exponential~\cite{ECHLPT92}. However, the computational complexity of the problem has not been made explicit\footnote{The implicit bound for the complexity of the word problem in 3-manifolds seems to be at least triply exponential. This follows from the exponential lower bounds for the isoperimetric function of a presentation of the fundamental group of a $3$-manifold~\cite[Theorem 8.1.3]{ECHLPT92}. Then the the trivial algorithm for solving the word problem~\cite[Theorem 2.2.5, Lemma 2.2.4]{ECHLPT92}, is triply exponential in the worst case. See also Aschenbrenner et al.~\cite{afw-dp3mf-15}.}. For decidability questions regarding 3-manifolds see, e.g., Aschenbrenner et al.~\cite[Section~4.1]{afw-dp3mf-15,AFW15}.

\subsection{Our result}

We consider a special case of the contractibility problem in 3-manifolds, namely, the case where the 3-manifold has non-empty boundary and the input curve lies on the \emph{boundary} of the 3-manifold.  Formally, the problem we consider is the following.

\begin{problem}
\BdContract$(m,n,t)$:\\
\Input: A triangulation~$T$ of a 3-manifold~$M$ made of $t$ tetrahedra; a general position PL closed curve $c$ on the boundary of~$M$ made of $n$ segments and having $m$ self-crossings.\\
\Output: The answer to the question ``Is $c$ contractible in~$M$?''.  If the answer is positive, and $c$ is non-contractible in~$\partial M$, the output also includes some simple curve in~$\partial M$ that is non-contractible in~$\partial M$ but bounds an embedded disk in~$M$.
\end{problem}

The \emph{decision problem associated to}~\BdContract{} is similar, except that the output is simply ``yes'' or ``no'' (without the possible curve in the output).

We remark that the size of the input of our algorithm is $O(n+t)$.  With our choice of representation of~$c$, as a PL curve, the self-crossings of~$c$ are not necessarily vertices of~$c$, so we always have $m=O(n^2)$, and we may have $m=\Theta(n^2)$.  However, $m$ is a natural parameter of the problem, so we choose to take it into account in the complexity analysis.

\begin{Theorem}\label{T:main}
  The \BdContract$(m,n,t)$ problem admits an algorithm with polynomial space complexity, and with time complexity $2^{O((m+n+t)^2)}$.
\end{Theorem}

Ignoring the possible curve output by the algorithm, this gives us an algorithm for deciding contractibility of a curve in~$M$, restricted to the case where the input curves lies on the boundary of~$M$.  In addition to being explicit and relatively simple, it has a running-time that is a considerable improvement over existing implicit bounds for the general contractibility problem for closed curves in a 3-manifold.

While the complexity of our algorithm is high, we remark that getting a polynomial complexity would result in a polynomial-time algorithm for the unknot problem (see below), the existence of which has been a longstanding open problem.  Moreover, our algorithm does not depend on the Geometrization Conjecture.

The proof of correctness builds closely upon the proof of the Loop Theorem by Papakyriacopoulos, the statement of which we recall here:
\begin{Theorem}[Loop Theorem~\cite{Pap57}]\label{theorem:loop1}
  Let $M$ be a $3$-manifold. Suppose that there is a closed curve that is non-contractible in $\partial M$ and contractible in~$M$.  Then there is a \emph{simple} closed curve that is non-contractible in~$\partial M$ and bounds an embedded disk in~$M$.
\end{Theorem}

Thus, our algorithm is naturally a computational version of the Loop Theorem:  If the input curve~$c$ is non-contractible in~$\partial M$ and contractible in~$M$, it computes a simple closed curve that is non-contractible in~$\partial M$ and bounds an embedded disk in~$M$.  Recall that deciding contractibility of curves on surfaces is easy in linear time~\cite{lr-hts-12,ew-tcsr-13}, so that it is possible to check whether $c$ is contractible in~$\partial M$ easily (although we will not need to do this in our algorithm: if the input curve~$c$ turns out to be contractible in~$\partial M$, the algorithm may return no curve, or may return a simple curve that is non-contractible in~$\partial M$ and bounds an embedded disk in~$M$).

By refining the analysis, we can obtain some related results, which we briefly indicate here, and which are detailed in the last section of this article.  

First, in a positive instance of \BdContract, we can, in the same amount of time, construct a singular (non-embedded) PL disk that bounds~$c$; it has the same space complexity as the running-time of the algorithm, see Theorem \ref{T:maindisk}. 

Secondly, the algorithm can in fact express the homotopy class of~$c$, in a tubular neighborhood~$F$ of its image, as an element in the normal subgroup generated by the homotopy classes, in~$F$, of a collection of closed simple curves~$C$, such that all curves in~$C$ are contractible in~$M$, see Theorem~\ref{T:mainformula}. In other words, the algorithm can output all possible simple curves whose existence is guaranteed by the Loop Theorem for~$c$, together with an expression writing $c$ as a product of their conjugates.

Finally, we prove that the \BdContract{} problem is in the complexity class~NP in two specific cases: (i) if the number~$m$ of self-intersections of the curve is bounded by a constant, or more generally by the logarithm of the input size (Theorem \ref{T:NP}); or (ii) if the boundary of~$M$ is a torus (Theorem \ref{theorem:torus})---this case is relevant because of the role of manifolds with torus boundaries in JSJ-decompositions~\cite{JaT95}.

\subsection{Related work}

\paragraph{More on the Loop Theorem.}
Historically, the Loop Theorem originates from Dehn's famous Lemma:  
\begin{Lemma}[Dehn's lemma~\cite{Deh10}]
  If a simple closed PL curve on the boundary of a 3-manifold bounds a PL disk, then it bounds an embedded PL disk.
\end{Lemma}
The first correct proof of the Loop Theorem was given by Papakyriakopoulos~\cite{Pap57} using the theory of covering spaces. Later, Shapiro and Whitehead~\cite{ShW58} presented a simpler proof of an extension of the Dehn's Lemma. This proof is the basis for the next extension by Stallings \cite{Sta60}. We heavily use the proof of this extended version, which can be found also in Rolfsen~\cite{Rol76}; see also Hempel~\cite{Hem04} and Shapiro and Whitehead~\cite{ShW58}. A proof of the Loop Theorem by quite a different approach is given by Aitchison and Rubinstein~\cite{ar-ldllt-04}.

\paragraph{The unknot problem.}
This is one of the most central problems in 3-dimensional algorithmic topology.  Formally, it is defined as follows.

\begin{problem}
\Unknot$(n)$:\\
\Input: A general position simple PL closed curve $c$ in~$\Rspace^3$, made of $n$ segments.\\
\Question: Is $c$ the unknot, namely, is~$c$ ambient isotopic to a circle in~$\Rspace^3$?
\end{problem}

It has been shown to be decidable by Haken~\cite{h-tnik-61}.  More recently, Hass et al.~\cite{HLP99} have proved that the problem is in NP.  In the very recent years, Kuperberg has shown that it is in co-NP assuming the generalized Riemann hypothesis~\cite{k-knmg-14}, and then Lackenby has announced that it is in co-NP unconditionally~\cite{l-ecktn-16}.  However, whether the problem is polynomial-time solvable remains a challenging open problem.

As it turns out, the unknot problem can be recast as a special case of the \BdContract{} problem, in which the curve~$c$ is \emph{simple}.  Indeed, Hass et al.~\cite{HLP99} solve the \Unknot{} problem by noting that the input closed curve~$c$ is unknotted if and only if, on the 3-manifold~$M$ obtained from~$\Rspace^3$ by removing a torus neighborhood of~$c$, a certain simple closed curve~$c'$ ``parallel'' to the input curve bounds an embedded disk in~$M$, or, equivalently by Dehn's lemma, is contractible in~$M$.  This reduction from \Unknot{} to \BdContract{} takes polynomial time, which implies, as announced above, that any polynomial-time algorithm for \BdContract{} would, in particular, yield a polynomial-time algorithm for \Unknot{}.

To provide a polynomial-size certificate, Hass et al.~\cite{HLP99} use normal surface theory developed by Haken~\cite{h-tnik-61} together with tools by Schubert~\cite{Sch61} and Jaco and Tollefson~\cite{JaT95}, which provides a compact encoding of some ``nice'' embedded surfaces in triangulated 3-manifolds; it turns out that a disk~$D$ as above can be represented by a normal surface of polynomial complexity.  This leads to an exponential-time algorithm for the contractibility problem for \emph{simple} curves on the boundary of a 3-manifold.  Our result implies an algorithm with a similar running time for the more general case of \emph{possibly non-simple} closed curves on the boundary of a 3-manifold.

Agol et al.~\cite{AHT06} extend significantly the results of Hass et al.~\cite{HLP99} to prove that for an arbitrary knot in a given orientable $3$-manifold, deciding if the knot bounds an embedded orientable surface of genus less than an input integer~$g$ is NP-complete.  The unknot problem is the same problem, restricted to the case where $g=0$ and where the 3-manifold is the 3-sphere.  The smallest integer~$g$ satisfying this property is the \emph{genus} of the knot.  Our algorithm relies on the above-mentioned algorithms from normal surface theory~\cite{Sch61,HLP99,AHT06} to decide contractibility of simple curves.

\paragraph{Waldhausen's algorithm.}
Even for curves in the interior of the 3-manifold, a special case for which a contractibility algorithm existed before the Geometrization Conjecture was proved is when the input 3-manifold is known to be sufficiently large and irreducible. This class includes the compact submanifolds of the 3-sphere. An algorithm for deciding contractibility in this case was given by Waldhausen~\cite{Wal68}. This algorithm uses Haken's decomposition of sufficiently large manifolds. The complexity of this algorithm has not been analyzed, however, it seems to be much more ineffective than the natural algorithm that computes in the fundamental group of the 3-manifold, using the isoperimetric bound whose existence is guaranteed using geometrization. See the complexity calculations by Haken~\cite{Hak62}.

\paragraph{Immersed normal surfaces.}
A natural idea for solving the contractibility problem of a non-simple curve~$c$ is to look for a singular (not necessarily embedded) disk bounded by~$c$.  However, by design, normal surface theory cannot handle non-embedded surfaces.  Recently, Burton et al.~\cite{bcm-cins-15} have studied a natural generalization of normal surfaces that could in principle handle surfaces that are not necessarily embedded.  However, they prove that it is NP-hard to decide whether such a relaxed normal surface corresponds to an \emph{immersed} surface, or equivalently whether it corresponds to a surface without a branchpoint. In any case, there is no bound known on the complexity of a singular disk bounding a non-simple curve on the boundary, other than those given by the implicit bounds on the general word problem for $3$-manifolds.


\subsection{Outline}

The rest of the article is organized as follows. After some preliminaries in Section~\ref{section:prelim}, we discuss the case where the input curve is simple (Section~\ref{section:simplecase}), and the cut-and-paste procedure on singular disks (Section~\ref{section:singularity}).  In Section~\ref{section:toyproblem}, we introduce another special case and show how this simpler case can be solved.  Sections \ref{section:tower} to~\ref{S:general} present our general algorithm and the theory behind its correctness.  Finally, in Section~\ref{section:extensions}, we discuss some extensions and special cases.

\section{Preliminaries and Notations}\label{section:prelim}

\subsection{Complexity theory}\label{S:NP}

We briefly introduce, somewhat informally, standard notions of complexity theory that we will need; see, e.g., Arora and Barak~\cite{ab-ccma-09} for more information.  In general, the efficiency of an algorithm is evaluated by its \emphdef{(time) complexity}, or \emphdef{running time}, which is an (asymptotic) function of the input size: It maps each input size to the largest number of standard operations that the algorithm will perform, over all inputs of size~$n$.  However, depending on the problem, it is often useful to introduce parameters of the problem others than the input size, and to express the complexity of the problem in terms of these parameters also.  As an example, in the problem \BdContract{}, the input size is $\Theta(n+t)$, but the parameter~$m$ is used also.  The \emphdef{time complexity} of a problem is the smallest time complexity of an algorithm solving it.  Similarly, the \emphdef{space complexity} of an algorithm is a function that maps each input size (or set of parameters) to the largest size of the computation space needed by the algorithm, over all inputs of that size (or with these parameters). The space complexity of a problem is defined analogously. We remark that the space needed to store a positive integer~$n$ is $\Theta(\log_2n)$, its \emph{bit complexity}, and that the standard arithmetic operations on numbers take time proportional to their bit complexities.

Here are some standard complexity classes for \emphdef{decision problems}, for which the answer is either ``yes'' or ``no'':
\begin{itemize}
    \item \emphdef{P} is the set of decision problems whose time complexity is upper bounded by a polynomial in the input size;
    \item \emphdef{NP} is, intuitively, the set of decision problems for which the ``yes'' instances can be ``certified'' to be indeed ``yes'' instances efficiently.  More formally, this is the set of decision problems for which there exists a \emphdef{verifier}, a polynomial-time algorithm~$V$ that takes as input the instance and some additional data of polynomial size, the \emphdef{certificate}, such that an instance~$i$ is a ``yes'' instance if and only if there exists some choice of certificate~$c$ such that $V$, when given as input $i$ and~$c$, returns ``yes'';
    \item \emphdef{co-NP} is the set of decision problems whose \emph{negation} is in NP;
    \item \emphdef{PSPACE} is the set of decision problems whose space complexity is upper bounded by a polynomial in the input size;
    \item \emphdef{EXPTIME} is the set of decision problems whose time complexity is upper bounded by an exponential function in a polynomial of the input size.
\end{itemize}
We have the inclusions P$\subseteq$NP$\subseteq$PSPACE$\subseteq$EXPTIME (and similarly with NP replaced with co-NP):
\begin{itemize}
    \item P$\subseteq$NP$\cap$co-NP because if a problem is in~P, we can define the verifier to run the polynomial-time algorithm, ignoring the certificate;
    \item NP$\cup$co-NP$\subseteq$PSPACE.  To prove NP$\subseteq$PSPACE, use the PSPACE algorithm that runs the NP-verifier over all possible certificates of polynomial size, and returns ``yes'' if and only if, for at least one certificate, the verifier returned ``yes''.  Of course, a symmetric argument holds for co-NP;
    \item PSPACE$\subseteq$EXPTIME because no algorithm running in polynomial space can be in the same state at two different times (otherwise, it would run forever), and the number of possible states of an algorithm running in polynomial space is exponential.
\end{itemize}
None of the above inclusions is known to be strict, except between the two extremes: P$\subsetneq$EXPTIME.

Thus, our result indicates that the decision problem associated to \BdContract{} lies in the complexity class~PSPACE.

\subsection{Topological notation}

We denote the standard unit disk in $\Rspace^2$ by \emphdef{$\bm{D^2}$}, and its boundary by~\emphdef{$\bm{S^1}$}.  We denote the homotopy class of a curve $c$ in $M$ by \emphdef{$\bm{[c]}$} and the homotopy class of $c$ in $\partial M$ by \emphdef{$\bm{[c]_\partial}$}. The basepoint for the fundamental groups is always a fixed point of $\partial M$.  In many cases, we are not explicit about the basepoints, because we are mainly concerned with contractibility of loops.  In general, we use subscripts to record the space in which a homotopy class is computed. A map of a pair \emphdef{$f:\bm{(D^2, \partial D^2) \rightarrow (M, \partial M)}$} is a map $f:D^2 \rightarrow M$ such that $f(\partial D^2) \subseteq\partial M$ and $f(D^2) \cap \partial M = f(\partial D^2)$. We remark that this latter requirement is not standard.  For standard topological concepts on homotopy and covering space, see, e.g., Hatcher~\cite{Hat01}.  

\subsection{Curves on triangulated 3-manifolds}


A \emphdef{triangulation} $T$ of a 3-manifold~$M$ is the combinatorial data of a number~$t$ of tetrahedra, with the information of a gluing between some pairs of their faces (i.e., identification of triangles), such that the resulting topological space, after gluing, is homeomorphic to~$M$.  (In particular, without loss of generality, $T$ could be chosen as a 3-dimensional simplicial complex, although generalized triangulations, which allow the same vertex, edge, or triangle to be incident several times to a given tetrahedron, are also admissible.)  Checking whether such a gluing of tetrahedra actually defines a 3-manifold with boundary is easy and standard, see, e.g., \cite[Proof of Theorem~2]{AHT06}; one needs to check that (i) each triangle is incident to at most two tetrahedra, (ii) no edge is identified with itself in reverse, (iii) the tetrahedra incident to a given vertex are ``sphere-like'' or ``disk-like'' (technically, the link of each vertex is a 2-dimensional sphere or a 2-dimensional disk).

We assume that our input 3-manifold~$M$ has non-empty boundary, which thus forms a surface (or 2-manifold)~$\partial M$.  Without loss of generality, we assume that the given curve $c$ is a piecewise-linear (PL) map in general position on the triangulated surface $\partial M$, by which we mean that it has finitely many self-intersection points, each of which is a double point where the intersection is transverse.  We represent~$c$ as a PL curve in~$\partial M$, in a way that the set of vertices of this PL curve includes the intersections of~$c$ with the edges of the triangulation of~$\partial M$, but not necessarily the self-intersections of~$c$.  In more detail, this means that $c$ is stored as an ordered sequence of its vertices, where a vertex lying in a triangle~$\tau$ of~$\partial M$ is described by its barycentric coordinates in~$\tau$.

\section{The Case of Simple Curves}\label{section:simplecase}

We begin by explaining the algorithm for contractibility in the case where the input curve (lies on the boundary of the 3-manifold and) is simple. We will use this algorithm for simple curves as a subroutine for the general case.

\begin{Proposition}\label{P:simple}
  The \BdContract$(0,n,t)$ problem, that is, the case where $c$ is simple ($m=0$), admits an algorithm with polynomial space complexity, and with time complexity $2^{O((n+t)^2)}$.  Moreover, the associated decision problem lies in~NP.
\end{Proposition}
\begin{proof}
  This essentially follows from previous work on algorithms using normal surface theory, and we will use some of the concepts of this theory, which are described in other articles containing the main ideas used below~\cite{HLP99,AHT06}.
  
  Let $T$ be the triangulation of the manifold~$M$.  We first compute, in polynomial time, another triangulation~$T'$ of~$M$ of complexity $O(n+t)$ such that the input curve~$c$ belongs to the 1-skeleton of~$T'$ (namely, the union of its vertices and edges of~$T'$).  For this purpose, let $\partial T'$ be a triangulation of~$\partial M$ that is a subdivision of~$T\cap\partial M$ containing the the image of~$c$ in its 1-skeleton; we can easily compute such a triangulation of complexity $O(n+t)$.  For each triangle~$\tau$ of~$T\cap\partial M$ belonging to a tetrahedron~$\sigma$ of~$T$, we subdivide~$\sigma$ by coning $\sigma\cap \partial T'$ from the vertex of~$\sigma$ not on~$\tau$.  Finally, for every tetrahedron~$\sigma$ of~$T$ intersecting~$\partial M$ along an edge split into, say, $k$ subedges in~$\partial T'$, we can easily subdivide~$\sigma$ into $k$ tetrahedra such that these subedges appear as edges of these tetrahedra.  Let $T'$ be the resulting triangulation; it satisfies the announced properties.
  
  By Dehn's lemma, if $c$ is contractible in~$M$, it bounds an embedded disk in~$M$.  By a result by Schubert~\cite[Section~5.5]{Sch61}, it then bounds a fundamental normal disk (with respect to the triangulation~$T'$).  By Hass et al.~\cite[Lemma~6.1(2)]{HLP99}, such a fundamental normal disk has size $O((n+t)^2)$ (in terms of number of bits needed to store it).  After these considerations, the algorithm for the \BdContract{} problem restricted to curves~$c$ that are simple is essentially described in Hass et al.~\cite[Section~8, \emph{Haken unknottedness algorithm}]{HLP99}.  It enumerates all normal surfaces with size $O((n+t)^2)$ and, for each of them, determines whether it is a disk bounding~$c$ (namely, whether it is connected, which can be done using the orbit counting algorithm of Agol et al.~\cite[Corollary~14]{AHT06}; and whether it has Euler characteristic one and bounds~$c$, which can be done easily using the normal coordinates).  If such a disk is found, it returns ``yes''; otherwise, it returns ``no''.  Moreover, if the answer is ``yes'', we determine whether $c$ is contractible in~$\partial M$; since $c$ is simple, it is equivalent to deciding whether $c$ bounds a disk in~$\partial M$, which is easy to do in linear time, by computing the Euler characteristic of the connected components of the surface $\partial M$ cut along~$c$.  If $c$ is non-contractible in~$\partial M$, then we also return it.  The algorithm is clearly correct and runs in $2^{O((n+t)^2)}$ time.
  
  There remains to prove that the decision problem associated to \BdContract{} is in NP.  The verification algorithm mentioned above for fundamental normal coordinates runs in polynomial time. This is indeed the reason for incorporating the orbit counting algorithm. See Agol et al.~\cite{AHT06}.
\end{proof}

  \emph{Remark.} Note that in the proof of Proposition \ref{P:simple} we have used the normal surface theory in the sense of Haken~\cite{Hak62} and Schubert~\cite{Sch61}. This setting seems more appropriate for the general problem of contractibility of simple curves than the now standard normal surface theory. In the former, one has different types of 3-dimensional cells and in the latter one works only with tetrahedra.

  \emph{Remark.} As an alternative to the above proof, we could use only the result by Agol et al.~\cite{AHT06}, which states that the problem of deciding if a simple closed curve in the interior of an orientable 3-manifold bounds an embedded surface of genus at most~$g$ is in NP. For this purpose, we could attach a collar to each boundary component of~$M$, obtaining a manifold~$M'$ containing~$c$ in its interior. We claim that $c$ is a knot of genus zero in $M'$ if and only if $c$ is contractible in $M$.  Indeed, if $c$ is a knot of genus zero in~$M'$, it is contractible in~$M'$, and thus also in~$M$ because $M'$ deformation retracts to~$M$.  Conversely, by Dehn's lemma, if $c$ is contractible in~$M$, then it bounds an embedded disk in~$M$, and thus also in~$M'$.  This proves the claim.  It thus suffices to decide whether $c$ has genus zero in~$M'$.  It is not difficult to build a triangulation of~$M'$ with complexity $O(n+t)$, and we can then use the NP proof (and corresponding algorithm) by Agol et al.~\cite{AHT06} directly.  That paper~\cite{AHT06} is presented in the case where the 3-manifold is orientable, but the results we use extend immediately to non-orientable ones by passing to the orientable double cover, which we can also construct algorithmically.

\section{Removing Singularities by Cut and Paste}\label{section:singularity}

We present procedures to replace some singular disks with others that have fewer singularities.  The ideas used in this section are instrumental in the proof of the Loop Theorem; see, e.g., Hempel~\cite[Chapter~4]{Hem04}.  However, additional details are needed because we do not only care about the existence of some curves, but we also need to be able to compute them.

\subsection{Smoothings of a curve in a $2$-manifold}

Let $c$ be a general position PL curve on~$\partial M$, and let $y \in c(S^1)$ be a self-crossing point of~$c$.

There are two natural ways to modify the curve $c$ in a small neighborhood of $c^{-1}(y)$, while staying on $\partial M$, such that the self-intersection $y$ is removed. These are depicted in Figure \ref{fig:smoothings}. Observe that one of the possibilities results in two curves and the other in a single curve. These operations are called the \emphdef{smoothings} of $c$ at $y$.

\myfigure{%
\begin{figure}
\centering
 \includegraphics[scale=0.25]{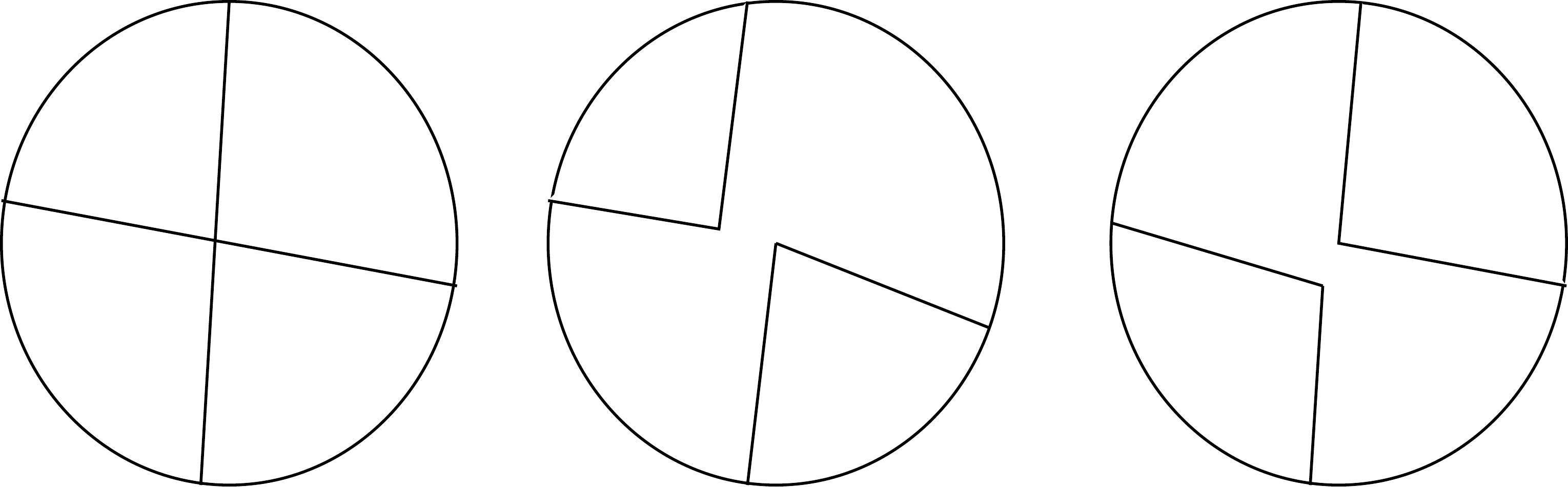}\caption{the two possible smoothings of a self-intersection.}
 \label{fig:smoothings}
\end{figure}
}

\subsection{Good singular disks}

A \emphdef{singular~disk} is a map $f\colon (D^2, \partial D^2)\to (M, \partial M)$.  In this article, we restrict ourselves to singular disks that are PL and in general position.  This does not restrict generality; indeed, if a PL curve~$c$ in~$\partial M$ is contractible, then the contraction can be taken to be PL and in general position, so $c$ bounds a PL singular disk in general position.  A \emphdef{singular point} of~$f$ is a point of~$M$ that is the image, under~$f$, of at least two points in~$D^2$.

We say that the singular disk~$f$ is \emphdef{good} if every singular point of~$f$ is the image of exactly two points in~$D^2$, and moreover the set of singular points is the disjoint union, in~$M$, of finitely many \emphdef{closed double curves}, i.e., simple closed curves, and \emphdef{double arcs}, i.e., simple paths in~$M$ whose endpoints have preimages in~$\partial D^2$.  Not every singular disk in general position is good; for example, a good singular disk contains no triple points (isolated points with three preimages in~$D^2$) or branchpoints (which have one preimage, but are limits of singular points---such branchpoints are excluded because, by definition, the set of singular points of a good singular disk~$f$ has to be closed).

Every contractible closed curve in~$\partial M$ bounds a singular disk.  However, not every contractible closed curve in~$\partial M$ bounds a good singular disk; for example, no contractible closed curve in~$\partial M$ with an odd number of self-intersections can bound a good singular disk.

\subsection{Eliminating closed double curves}

Eliminating closed double curves in a good singular disk is easy:
\begin{Lemma}\label{lemma:closeddoublecurve}
  Let $M$ be a $3$-manifold and $f:(D^2,\partial D^2) \rightarrow (M, \partial M)$ be a good singular disk.  Then there is a good singular disk $g:(D^2,\partial D^2)\rightarrow (M,\partial M)$ such that $g|_{\partial D^2}=f|_{\partial D^2}$ and $f$ has no closed double curve.
\end{Lemma}
\begin{proof}
  This directly follows from Hempel~\cite[Lemma~4.6]{Hem04}.  Here is an overview of the proof (we deliberately omit some details, especially in the case where $M$ is non-orientable).  A closed double curve in~$f$ corresponds to two disjoint closed curves $d'$ and~$d''$ in the interior of~$D^2$, bounding disks $D'$ and~$D''$ in~$D^2$.  There are two cases depending on whether $d'$ and~$d''$ are nested.  If they are nested, without loss of generality we can assume that $d'$ is the exterior closed curve.  In~$f$, we can replace $f|_{D'}$ with $f|_{D''}$ (suitably reparameterized).  If they are not nested, then one can replace $f|_{D''}$ with $f|_{D'}$ (suitably reparameterized) and similarly $f|_{D'}$ with $f|_{D''}$, and finally slightly perturb the resulting map in a suitable way.  This removes at least one closed double curve; iterating, we remove all closed double curves.
\end{proof}

\subsection{Eliminating double arcs}

Eliminating double arcs is more complicated, because such arcs intersect the boundary of the singular disk.  We first need a definition.  Let $c$ be a closed curve on~$\partial M$, and $u$ and~$v$ be two self-intersection points of~$c$.  When walking along~$c$, the points $u$ and~$v$ are encountered twice, either in the order $uuvv$ (\emphdef{non-interlaced~case}) or in the order $uvuv$ (\emphdef{interlaced~case})---these are the only possibilities up to cyclic permutation.

Below, we describe two ways to eliminate a double arc, depending on whether the endpoints of the double arc are interlaced or not along~$c$.  These arguments are essentially from Hempel \cite[Lemma~4.7]{Hem04}.  

First, let us consider the easier non-interlaced case.
\begin{Lemma}\label{lemma:doublearc-noninter}
  Let $M$ be a $3$-manifold, and let $c:S^1\to\partial M$ be a closed curve on~$\partial M$.  Let $F$ be a tubular neighborhood of the image of~$c$.  Let $u$ and~$v$ be distinct self-crossings of~$c$ that are non-interlaced.  Then we can compute two closed curves $c'$ and~$c''$ in~$F$, each resulting from smoothing $c$ in~$\partial M$ at~$u$ and~$v$, such that:
  \begin{itemize}
\item $[c]_F$ is in the normal subgroup generated by $[c']_F$ and $[c'']_F$;
\item if $c$ bounds a good singular disk~$f:D^2\to M$ such that $u$ and~$v$ are endpoints of a double arc~$a$, then $c'$ and~$c''$ bound good singular disks $f'$ and~$f''$ in~$M$, whose sets of double arcs are included in those of~$f$, but do not contain~$a$.
\end{itemize}
\end{Lemma}
\myfigure{
\begin{figure}
 \centering{\def\svgwidth{.8\linewidth}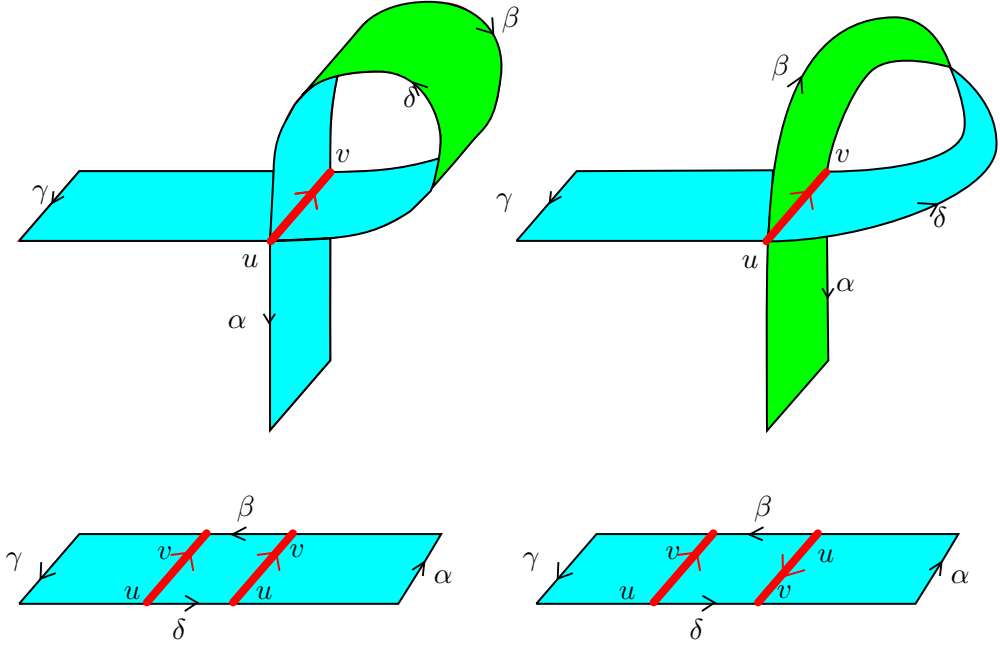}\caption{Two self-crossing disks (top), corresponding to two different cases for the preimages of the arcs in the disk~$D^2$ (bottom). The non-interlaced case (Lemma~\ref{lemma:doublearc-noninter}) is represented on the left, and the interlaced case (Lemma~\ref{lemma:doublearc-inter}) is represented on the right.}
 \label{fig:double-arcs-two-types}
\end{figure}
}
\myfigure{
\begin{figure}\small
 \centering
 {\def\svgwidth{.3\linewidth}
\begingroup%
  \makeatletter%
  \providecommand\color[2][]{%
    \errmessage{(Inkscape) Color is used for the text in Inkscape, but the package 'color.sty' is not loaded}%
    \renewcommand\color[2][]{}%
  }%
  \providecommand\transparent[1]{%
    \errmessage{(Inkscape) Transparency is used (non-zero) for the text in Inkscape, but the package 'transparent.sty' is not loaded}%
    \renewcommand\transparent[1]{}%
  }%
  \providecommand\rotatebox[2]{#2}%
  \newcommand*\fsize{\dimexpr\f@size pt\relax}%
  \newcommand*\lineheight[1]{\fontsize{\fsize}{#1\fsize}\selectfont}%
  \ifx\svgwidth\undefined%
    \setlength{\unitlength}{172.28586095bp}%
    \ifx\svgscale\undefined%
      \relax%
    \else%
      \setlength{\unitlength}{\unitlength * \real{\svgscale}}%
    \fi%
  \else%
    \setlength{\unitlength}{\svgwidth}%
  \fi%
  \global\let\svgwidth\undefined%
  \global\let\svgscale\undefined%
  \makeatother%
  \begin{picture}(1,0.84937248)%
    \lineheight{1}%
    \setlength\tabcolsep{0pt}%
    \put(0,0){\includegraphics[width=\unitlength,page=1]{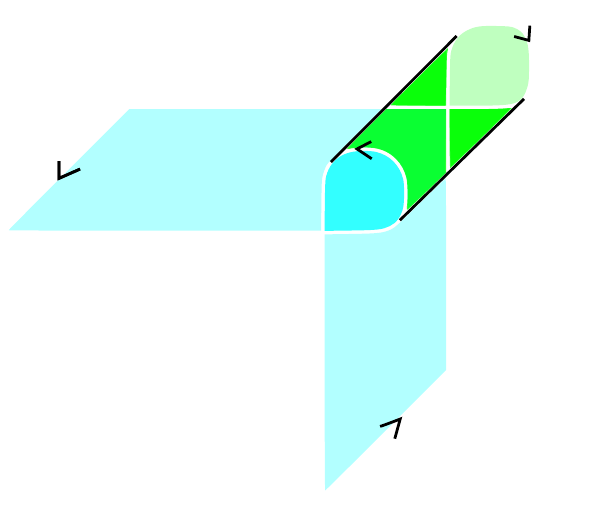}}%
    \put(0.07963047,0.59630339){\color[rgb]{0,0,0}\makebox(0,0)[lt]{\lineheight{0}\smash{\begin{tabular}[t]{l}$\gamma$\end{tabular}}}}%
    \put(0.62878107,0.60697447){\color[rgb]{0,0,0}\makebox(0,0)[lt]{\lineheight{0}\smash{\begin{tabular}[t]{l}$\delta$\end{tabular}}}}%
    \put(0.67286673,0.1088006){\color[rgb]{0,0,0}\makebox(0,0)[lt]{\lineheight{0}\smash{\begin{tabular}[t]{l}$\alpha$\end{tabular}}}}%
    \put(0.86822977,0.822913){\color[rgb]{0,0,0}\makebox(0,0)[lt]{\lineheight{0}\smash{\begin{tabular}[t]{l}$\beta$\end{tabular}}}}%
    \put(0,0){\includegraphics[width=\unitlength,page=2]{simplify1-disk.pdf}}%
    \put(0.45694356,0.40217593){\color[rgb]{0,0,0}\makebox(0,0)[lt]{\lineheight{1.25}\smash{\begin{tabular}[t]{l}$u$\end{tabular}}}}%
    \put(0.74860998,0.68078266){\color[rgb]{0,0,0}\makebox(0,0)[lt]{\lineheight{1.25}\smash{\begin{tabular}[t]{l}$v$\end{tabular}}}}%
  \end{picture}%
\endgroup%
}{\def\svgwidth{.3\linewidth}
\begingroup%
  \makeatletter%
  \providecommand\color[2][]{%
    \errmessage{(Inkscape) Color is used for the text in Inkscape, but the package 'color.sty' is not loaded}%
    \renewcommand\color[2][]{}%
  }%
  \providecommand\transparent[1]{%
    \errmessage{(Inkscape) Transparency is used (non-zero) for the text in Inkscape, but the package 'transparent.sty' is not loaded}%
    \renewcommand\transparent[1]{}%
  }%
  \providecommand\rotatebox[2]{#2}%
  \newcommand*\fsize{\dimexpr\f@size pt\relax}%
  \newcommand*\lineheight[1]{\fontsize{\fsize}{#1\fsize}\selectfont}%
  \ifx\svgwidth\undefined%
    \setlength{\unitlength}{143.14697525bp}%
    \ifx\svgscale\undefined%
      \relax%
    \else%
      \setlength{\unitlength}{\unitlength * \real{\svgscale}}%
    \fi%
  \else%
    \setlength{\unitlength}{\svgwidth}%
  \fi%
  \global\let\svgwidth\undefined%
  \global\let\svgscale\undefined%
  \makeatother%
  \begin{picture}(1,0.81118417)%
    \lineheight{1}%
    \setlength\tabcolsep{0pt}%
    \put(0,0){\includegraphics[width=\unitlength,page=1]{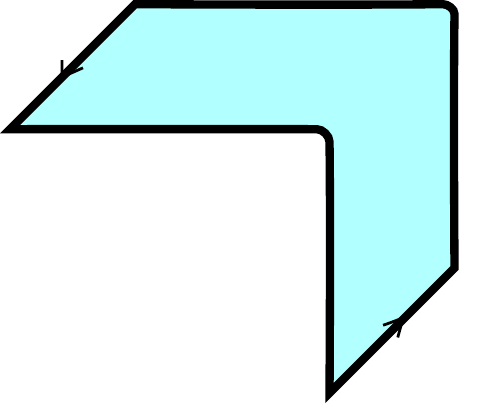}}%
    \put(0.10178796,0.71021472){\color[rgb]{0,0,0}\makebox(0,0)[lt]{\lineheight{0}\smash{\begin{tabular}[t]{l}$\gamma$\end{tabular}}}}%
    \put(0.81971245,0.10120892){\color[rgb]{0,0,0}\makebox(0,0)[lt]{\lineheight{0}\smash{\begin{tabular}[t]{l}$\alpha$\end{tabular}}}}%
  \end{picture}%
\endgroup%
}{\def\svgwidth{.3\linewidth}
\begingroup%
  \makeatletter%
  \providecommand\color[2][]{%
    \errmessage{(Inkscape) Color is used for the text in Inkscape, but the package 'color.sty' is not loaded}%
    \renewcommand\color[2][]{}%
  }%
  \providecommand\transparent[1]{%
    \errmessage{(Inkscape) Transparency is used (non-zero) for the text in Inkscape, but the package 'transparent.sty' is not loaded}%
    \renewcommand\transparent[1]{}%
  }%
  \providecommand\rotatebox[2]{#2}%
  \newcommand*\fsize{\dimexpr\f@size pt\relax}%
  \newcommand*\lineheight[1]{\fontsize{\fsize}{#1\fsize}\selectfont}%
  \ifx\svgwidth\undefined%
    \setlength{\unitlength}{177.17094395bp}%
    \ifx\svgscale\undefined%
      \relax%
    \else%
      \setlength{\unitlength}{\unitlength * \real{\svgscale}}%
    \fi%
  \else%
    \setlength{\unitlength}{\svgwidth}%
  \fi%
  \global\let\svgwidth\undefined%
  \global\let\svgscale\undefined%
  \makeatother%
  \begin{picture}(1,0.81342807)%
    \lineheight{1}%
    \setlength\tabcolsep{0pt}%
    \put(0,0){\includegraphics[width=\unitlength,page=1]{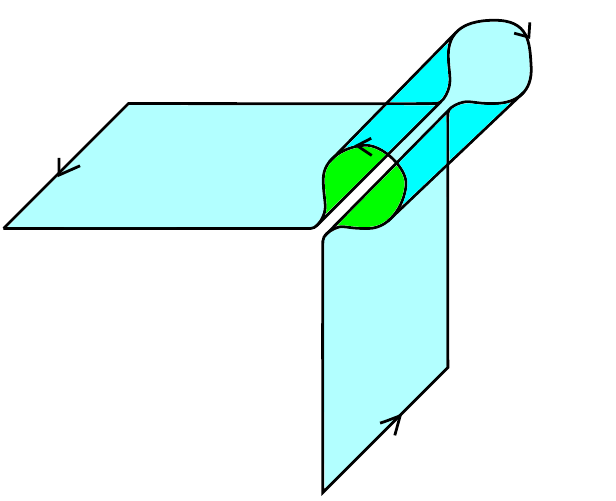}}%
    \put(0.07743485,0.57267768){\color[rgb]{0,0,0}\makebox(0,0)[lt]{\lineheight{0}\smash{\begin{tabular}[t]{l}$\gamma$\end{tabular}}}}%
    \put(0.61144387,0.58305453){\color[rgb]{0,0,0}\makebox(0,0)[lt]{\lineheight{0}\smash{\begin{tabular}[t]{l}$\delta$\end{tabular}}}}%
    \put(0.65794017,0.08965494){\color[rgb]{0,0,0}\makebox(0,0)[lt]{\lineheight{0}\smash{\begin{tabular}[t]{l}$\alpha$\end{tabular}}}}%
    \put(0.87186303,0.78769815){\color[rgb]{0,0,0}\makebox(0,0)[lt]{\lineheight{0}\smash{\begin{tabular}[t]{l}$\beta$\end{tabular}}}}%
    \put(0,0){\includegraphics[width=\unitlength,page=2]{simplify2-disk.pdf}}%
  \end{picture}%
\endgroup%
}
 \caption{Left: A self-crossing disk~$f$ as in Figure~\ref{fig:double-arcs-two-types}, left (i.e., the non-interlaced case).  Middle: the self-crossing disk~$f'$.  Right: the self-crossing disk~$f''$.}
 \label{fig:simplify-disk}
\end{figure}
}%
\begin{proof}
Let us first assume that $f$ is a good singular disk containing a double arc~$a$ with endpoints $u$ and~$v$, which are non-interlaced; see Figure~\ref{fig:double-arcs-two-types}(left).   Write $c$ as the concatenation of $\alpha$, $\beta$, $\gamma$, and~$\delta$ in this order, in such a way that $\alpha$ goes from~$u$ to~$v$, $\gamma$ goes from~$v$ to~$u$, and both $\beta$ and~$\delta$ are loops based at~$v$ and~$u$, respectively.  In two different ways, we can simplify~$f$ to remove the arc~$a$ by cutting the image of~$f$ open along~$a$ and then gluing back the pieces; see Figure~\ref{fig:simplify-disk}.  These two operations result in two good singular disks $f'$ and~$f''$, bounded by curves $c':=f'|_{\partial D^2}=\alpha\cdot\gamma$ and $c'':=f''|_{\partial D^2}=\alpha\cdot\beta\inv\cdot\gamma\cdot\delta\inv$.

The crucial point in the above construction is that the definition of $c'$ and~$c''$ does not depend on the singular disk~$f$, but only on~$c$, $u$, and~$v$.  Thus, let $c$, $u$, and~$v$ be as in the statement of the lemma (and let us not assume any more that $u$ and~$v$ are the endpoints of a double arc).  See Figure~\ref{fig:simplify-disk}.  As above, write $c$ as the concatenation of $\alpha$, $\beta$, $\gamma$, and~$\delta$ in this order, in such a way that $\alpha$ goes from~$u$ to~$v$, $\gamma$ goes from~$v$ to~$u$, and both $\beta$ and~$\delta$ are loops based at~$v$ and~$u$, respectively.  We define $c':=\alpha\cdot\gamma$ and $c'':=\alpha\cdot\beta\inv\cdot\gamma\cdot\delta\inv$.  These curves are obtained by smoothing~$c$ at $u$ and~$v$.  A trivial computation shows that
{\def\q#1{[#1]_F\cdot}%
\def\qq#1{[#1]_F}%
$\qq{c}=\q{c'}\q{\delta\inv}\q{c^{\prime\prime -1}}\q{c'}\qq{\delta}$}, which implies that $[c]_F$ is in the normal subgroup generated by $[c']_F$ and~$[c'']_F$.

Moreover, if $c$ turns out to bound a good singular disk such that $u$ and~$v$ are endpoints of a double arc~$a$, then the above construction shows that $c'$ and~$c''$ bound good singular disks in~$M$, whose sets of double arcs are included in those of~$f$, but do not contain the arc~$a$.
\end{proof}

Now, let us consider the interlaced case.  The statement is similar in spirit, but more complicated: There are two pairs of curves to consider.
\begin{Lemma}\label{lemma:doublearc-inter}
  Let $M$ be a $3$-manifold, and let $c:S^1\to\partial M$ be a closed curve on~$\partial M$.  Let $F$ be a tubular neighborhood of the image of~$c$.  Let $u$ and~$v$ be distinct self-crossings of~$c$ that are interlaced.  Then we can compute four closed curves $c'_i$ and~$c''_i$, $i=1,2$, in~$F$, each resulting from smoothing $c$ in~$\partial M$ at~$u$ and~$v$, such that:
  \begin{itemize}
\item for each $i=1,2$, $[c]_F$ is in the normal subgroup generated by $[c'_i]_F$ and $[c''_i]_F$;
\item if $c$ bounds a good singular disk~$f:D^2\to M$ such that $u$ and~$v$ are endpoints of a double arc~$a$, then for some $i\in\{1,2\}$, the curves $c'_i$ and~$c''_i$ bound good singular disks $f'$ and~$f''$ in~$M$, whose sets of double arcs are included in those of~$f$, but do not contain~$a$.
\end{itemize}
\end{Lemma}
\begin{proof}
  Let us first assume that $f$ is a good singular disk containing a double arc~$a$, whose endpoints are interlaced along~$c$; see Figure~\ref{fig:double-arcs-two-types}(right).   Write $c$ as the concatenation of $\alpha$, $\beta$, $\gamma$, and~$\delta$ in this order, where the endpoints of these four paths are $u$ and~$v$.  By construction, either $\alpha$ and~$a$ together bound a good singular disk whose image is included in the image of~$f$, or $\beta$ and~$a$ do; up to changing $\alpha$, $\beta$, $\gamma$, and~$\delta$, let us assume that $\alpha$ and~$a$ do.  As in the preceding proof, we can, in two different ways, simplify~$f$ to remove the arc~$a$ by cutting the image of~$f$ open along~$a$ and then gluing back the pieces.  These two operations result in two good singular disks $f'$ and~$f''$, bounded by curves $c':=f'|_{\partial D^2}=\alpha\cdot\gamma\inv$ and $c'':=f''|_{\partial D^2}=\alpha\cdot\delta\cdot\gamma\cdot\beta$.

  Now, let $c$, $u$, and~$v$ be as in the statement of the lemma (and let us not assume any more that $u$ and~$v$ are the endpoints of a double arc).  Write $c$ as the concatenation of $\alpha$, $\beta$, $\gamma$, and~$\delta$ in this order, where the endpoints of these four paths are $u$ and~$v$.  We define $c':=\alpha\cdot\gamma\inv$ and $c'':=\alpha\cdot\delta\cdot\gamma\cdot\beta$.  These curves are obtained by smoothing~$c$ at $u$ and~$v$.  A trivial computation shows that
{\def\q#1{[#1]_F\cdot}%
\def\qq#1{[#1]_F}%
$\qq{c}=\q{c'}\q{(\gamma\cdot\delta)\inv}\q{c^{\prime -1}}\q{c''}\qq{\gamma\cdot\delta}$}, which implies that $[c]_F$ is in the normal subgroup generated by $[c']_F$ and~$[c'']_F$.  

Assume moreover that $c$ bounds a good singular disk~$f$, and that $\alpha$ and~$a$ together bound a good singular disk whose image is included in the image of~$f$.  Then the above construction shows that $c'$ and~$c''$ bound good singular disks in~$M$, whose sets of double arcs are included in those of~$f$, but do not contain the arc~$a$.

The subtlety compared to the previous proof is that, given~$c$, written as the concatenation $\alpha\cdot\beta\cdot\gamma\cdot\delta$ where the endpoints of these four paths are $u$ and~$v$, and assuming it bounds a good singular disk~$f$, the algorithm cannot decide whether $\alpha$ and~$a$ (and thus also $\gamma$ and~$a$) bound a good singular disk whose image is included in the image of~$f$, or whether $\beta$ and~$a$ (and thus also $\delta$ and~$a$) do.  So, we have to introduce two pairs of curves, one for each case.  Specifically, define $c'_1:=\alpha\cdot\gamma\inv$ and $c''_1:=\alpha\cdot\delta\cdot\gamma\cdot\beta$ (the case described above, in which $\alpha$ and~$a$ together bound a singular disk whose image is included in that of~$f$), and similarly $c'_2:=\beta\cdot\delta\inv$ and $c''_2:=\beta\cdot\alpha\cdot\delta\cdot\gamma$ (corresponding to the case where $\beta$ and~$a$ together bound a singular disk whose image is included in that of~$f$). As a side note, it turns out that $c''_1$ and~$c''_2$ are identical, except for their basepoints; in the sequel, we do not exploit this coincidence.
\end{proof}

\section{Strongly Contractible Curves} \label{section:toyproblem}

Not every contractible closed curve in~$\partial M$ bounds a good singular disk in~$M$.  We call a closed curve on $\partial M$ \emphdef{strongly contractible} if it bounds a good singular disk.  The purpose of this section is to prove the following proposition.
\begin{Proposition}\label{P:toy}
  Under the hypotheses of Theorem~\ref{T:main}, there is an algorithm that in polynomial space and $2^{O((m+n+t)^2)}$ time outputs either ``(i)'' or ``(ii)'', such that:
\begin{itemize}
    \item if it outputs ``(i)'', then $c$ is contractible in~$M$;
    \item if it outputs ``(ii)'', then $c$ is not strongly contractible in~$M$.
\end{itemize}
Moreover, if it outputs~(i), and if $c$ is non-contractible in~$\partial M$, it also outputs a simple curve contractible in~$M$ and non-contractible in~$\partial M$.
\end{Proposition}
Note that these two possibilities are not exclusive from each other; if the input curve~$c$ is contractible but not strongly contractible, the algorithm can return either (i) or~(ii).  Roughly speaking, if it were true that every contractible closed curve on~$\partial M$ is strongly contractible, then the above proposition would give an algorithm for deciding contractibility.

Although we will not use this proposition directly in the rest of the paper, the arguments of the proof will be reused almost verbatim in a more refined context later (see Proposition~\ref{P:special}).

The proof uses the following lemma.
\begin{Lemma}\label{L:toy}
  Assume that $c$ is strongly contractible but is not simple.  Then there is a pair of (distinct) self-intersection points $u$ and~$v$ of~$c$ satisfying the following:
  \begin{itemize}
      \item if $u$ and~$v$ are non-interlaced along~$c$, then the curves $c'$ and~$c''$ obtained from smoothing~$c$ at $u$ and~$v$ (as in Lemma~\ref{lemma:doublearc-noninter}) are strongly contractible;
      \item if $u$ and~$v$ are interlaced along~$c$, then, for some $i\in\{1,2\}$, the curves $c'_i$ and~$c''_i$ obtained from smoothing~$c$ at $u$ and~$v$ (as in Lemma~\ref{lemma:doublearc-inter}) are strongly contractible.
  \end{itemize}
\end{Lemma}
\begin{proof}
  Since $c$ bounds a good singular disk but is not simple, it has at least two self-intersection points $u$ and~$v$ connected by a double arc.  Lemmas~\ref{lemma:doublearc-noninter} and \ref{lemma:doublearc-inter} immediately imply the result.
\end{proof}

\begin{proof}[Proof of Proposition~\ref{P:toy}]
  We note that the following algorithm is not the most efficient one could think of, but the same proof pattern will be used in the proof of Proposition~\ref{P:special}.%
  \global\long\outer\def\proofofprop#1#2{The algorithm is recursive.  If $c$ has no self-intersection point, we first determine whether $c$ is contractible in~$M$, using Proposition~\ref{P:simple}.  If it is the case, then we return~(i) ($c$ is contractible in~$M$), together with the curve~$c$ if output by the algorithm of Proposition~\ref{P:simple}; otherwise, we return~(ii) ($c$ is not #1); in both cases, this is correct.  If $c$ has a single self-intersection point, we return~(ii); this is also correct, because otherwise, $c$ would be #1{} and would thus have at least two self-intersection points (Lemma~\ref{#2}).

  Now, assume that $c$ has at least two self-intersection points.  We apply the following algorithm.
  \begin{enumerate}
      \item For every choice of distinct self-intersection points $u$ and~$v$ of~$c$, we do the following:
      \begin{enumerate}
      \item if $u$ and~$v$ are non-interlaced: compute the curves $c'$ and~$c''$ obtained from Lemma~\ref{lemma:doublearc-noninter}, and recursively run the algorithm on $c'$ and~$c''$.  If the outcomes for both $c'$ and~$c''$ is~(i), then return~(i).
      \item if $u$ and~$v$ are interlaced: compute the curves $c'_1$, $c''_1$, $c'_2$, and~$c''_2$ obtained from Lemma~\ref{lemma:doublearc-inter}, and recursively run the algorithm on these four curves.  If the outcome for both $c'_1$ and~$c''_1$, or for both $c'_2$ and~$c''_2$, is~(i), then return~(i).
      \end{enumerate}
      \item (Assuming the algorithm arrives at this point:) Return~(ii).
   \end{enumerate}
   We now prove correctness.  Assume that the algorithm returns~(i) in Step 1(a) or~1(b).  This is correct: indeed, if this occurs in Step~1(a), by Lemma~\ref{lemma:doublearc-noninter}, the homotopy class of $c$ lies in the normal subgroup generated by the homotopy classes of $c'$ and~$c''$, in~$\partial M$, and thus also in~$M$; but on the other hand, by~(i), $c'$ and~$c''$ are contractible in~$M$; so $c$ is indeed contractible in~$M$.  
   If this occurs in Step~1(b), by Lemma~\ref{lemma:doublearc-inter}, the same argument holds, in which $c'$ and~$c''$ are replaced with $c'_i$ and~$c''_i$ for some $i\in\{1,2\}$.
   
   Moreover, if the output of at least one of these two recursive calls contains a curve, we return one such curve.  By the induction hypothesis, the output curve satisfies the desired properties.  Finally, if no curve is output, this implies that both $c'$ and~$c''$ (or $c'_i$ and~$c''_i$, for some $i\in\{1,2\}$, in the interlaced case) are contractible in~$\partial M$, so $c$, the homotopy class of which lies in the normal subgroup generated by those of~$c'$ and~$c''$ (or $c'_i$ and~$c''_i$) in~$\partial M$, is also contractible in~$\partial M$; hence it is correct to output no curve.
 
  Otherwise, we return~(ii).  This is correct because otherwise, Lemma~\ref{L:toy} would imply that for some choice of $u$ and~$v$, both $c'$ and~$c''$ (or $c'_i$ and~$c''_i$) are #1, which we know is not true.

  The space needed by the algorithm is polynomial.  Indeed, the depth of the recursion tree is $m/2$, and each call takes polynomial space.  The running time of the algorithm can be bounded from above by a function~$\varphi(m)$ of the number~$m$ of self-intersections of~$c$.  Each of the closed curves at the leaves of the recursion tree has complexity $O(m+n+t)$.  We have $\varphi(0)=2^{O((m+n+t)^2)}$ by Proposition~\ref{P:simple} (because the curves at the leaves of the recursion tree can have complexity $O(m+n+t)$); we also have $\varphi(1)=O(1)$ and $\varphi(m)\le m(m-1)\cdot 4(\varphi(m-2)+O(m+n+t))$ for $m\ge2$.  This implies by induction $\varphi(m)\le m!2^{2m}\max\{\varphi(0),\varphi(1)\}+2^{2m}O((m+n+t)^3)$, and thus the result.}

  \proofofprop{strongly contractible}{L:toy}
\end{proof}
  
\section{Towers of Two-Sheeted Covering Spaces}\label{section:tower}

As we have seen in the previous section, if it were true that every contractible closed curve on~$\partial M$ is strongly contractible, we would have an algorithm to test such curves for contractibility.  It is not the case, however, and the proof of the Loop Theorem has to face this difficulty.  The way to circumvent it is via an iterative construction of two-sheeted covering spaces, a concept that originates from the classical proof of the Loop Theorem due to Shapiro and Whitehead~\cite{ShW58} and available in various sources~\cite{Hem04,Rol76,h-nb3mt-07}, which we reformulate now.

Let $f:D^2\to M$ be a singular disk.  
A \emphdef{tower} (of two-sheeted covering spaces) for~$f$ is defined by the following commutative diagram:

\begin{diagram}
M_0          & \lInto &V_0         &              &   \\ 
&\rdTo^{p_0} &        &\luTo^{f_0}(2,6) &              &   \\
M_{1}      & \lInto &V_{1}             &              &   \\
\vdots          &        &\vdots              &\luTo(2,4)    &   \\
M_{\ell-1}      & \lInto &V_{\ell-1}  &              &    \\ 
&\rdTo^{p_{\ell-1}}    &        &\luTo^{f_{\ell-1}}         &              & \\
M=M_\ell           & \lInto &V_\ell            &\lInto^{f_\ell=f}&D^2\\  
\end{diagram}
such that for each~$i$:
\begin{enumerate}
\item $M_i$ and $V_i$ are connected 3-manifolds with boundary;
\item the map $p_{i-1}:M_{i-1}\to V_i$ is a two-sheeted covering map;
\item $f_i$ maps $\partial D^2$ into~$\partial V_i$ and (after composition with the inclusion map $V_i\hookrightarrow M_i$) into~$\partial M_i$.

\end{enumerate}

We also say that this tower is a tower for the curve~$f|_{\partial D^2}$.

The following theorem follows from the classical proofs of the Loop Theorem~\cite{ShW58,Sta60}, and will be used in Section~\ref{S:general}:
\begin{Theorem}\label{theorem:essentialloop}
Under the hypotheses of Theorem~\ref{T:main}, assume that $c$ is contractible. Then there is a tower for~$c$ such that $\partial V_0$ is a disjoint union of spheres.
\end{Theorem}

\section{Virtually Strongly Contractible Curves}\label{S:special}

We say that a curve $c$ on~$\partial M$ is \emphdef{virtually strongly contractible} if there exists a tower for~$c$ such that $c$ lifts to a strongly contractible curve in~$\partial V_0$.  In particular, every virtually strongly contractible curve is contractible in~$M$, because it lifts to a contractible closed curve.

We now prove the following proposition, which is the same as Proposition~\ref{P:toy} with ``strongly contractible'' replaced by ``virtually strongly contractible'':
\begin{Proposition}\label{P:special}
  Under the hypotheses of Theorem~\ref{T:main}, there is an algorithm that in polynomial space and $2^{O((m+n+t)^2)}$ time outputs either ``(i)'' or ``(ii)'', such that:
\begin{itemize}
    \item if it outputs ``(i)'', then $c$ is contractible in~$M$;
    \item if it outputs ``(ii)'', then $c$ is not virtually strongly contractible.
\end{itemize}
Moreover, in case~(i), and if $c$ is non-contractible in~$\partial M$, it also outputs a simple curve contractible in~$M$ and non-contractible in~$\partial M$.
\end{Proposition}

The proof rests on the following lemma.
\begin{Lemma}\label{L:special}
  Assume that $c$ is strongly contractible but is not simple.  Then there is a pair of (distinct) self-intersection points $u$ and~$v$ of~$c$ satisfying the following:
  \begin{itemize}
      \item if $u$ and~$v$ are non-interlaced along~$c$, then the curves $c'$ and~$c''$ obtained from smoothing~$c$ at $u$ and~$v$ (as in Lemma~\ref{lemma:doublearc-noninter}) are virtually strongly contractible;
      \item if $u$ and~$v$ are interlaced along~$c$, then, for some $i\in\{1,2\}$, the curves $c'_i$ and~$c''_i$ obtained from smoothing~$c$ at $u$ and~$v$ (as in Lemma~\ref{lemma:doublearc-inter}) are virtually strongly contractible.
  \end{itemize}
\end{Lemma}
\begin{proof}
  Consider a tower for~$c$, with minimum possible height~$\ell$ among all choices of towers of~$c$ such that $f_0$ is a good singular disk.  By Lemma~\ref{lemma:closeddoublecurve}, we can assume that $f_0$ has no closed double curve.
    
  Assume that $f_0$ is an embedding.  This implies $\ell\ge1$, because $f$ is not an embedding due to the self-crossings of~$c$.  Moreover, because $f_0$ is an embedding and $p_0$ is a two-sheeted covering map, $f_1$ is a good singular disk (this is the key reason why two-sheeted covering spaces are used!).  So the tower starting at $V_1$, not $V_0$, is a tower of smaller height; this contradicts the choice of the tower.  Hence, $f_0$ is a good singular disk with at least one double curve, but no closed double curve (as shown in the previous paragraph), thus with at least one double arc.
    
  We apply Lemma~\ref{lemma:doublearc-noninter} or Lemma~\ref{lemma:doublearc-inter} in~$V_0$ to~$f_0$ and to the endpoints of this double arc, depending on whether these endpoints are interlaced along~$c$.  We obtain two closed curves obtained from smoothing $f_0|_{\partial D^2}$ at these endpoints, which bound good singular disks.  Let $c'$ and~$c''$ be the projections of these two closed curves to~$M$; they are also obtained from~$c$ by the smoothings described in Lemma~\ref{lemma:doublearc-noninter} or~\ref{lemma:doublearc-inter}, because the constructions of~$c'$ and~$c''$ are done in the same way regardless of whether they take place in~$M$ or in any space in the tower.  This implies that the tower for~$c$ is also a tower for $c'$ and~$c''$, which are thus also virtually strongly contractible.
\end{proof}

\begin{proof}[Proof of Proposition~\ref{P:special}]
  The algorithm is identical to that in the proof of Proposition~\ref{P:toy}. The proof is exactly the same, replacing everywhere ``strongly contractible'' with ``virtually strongly contractible'', and using Lemma~\ref{L:special} in place of Lemma~\ref{L:toy}.  We reproduce the proof for clarity.
  
  \proofofprop{virtually strongly contractible}{L:special}
\end{proof}
  
\section{Proof of Theorem~\ref{T:main}}\label{S:general}

Let $c$ be a closed curve on~$\partial M$ in general position, let \emphdef{$\bm{C}$} be the set of self-crossing points of~$c$, and let $X\subseteq C$.  Let \emphdef{$\bm{G_X}$} be the abstract graph, naturally drawn (with crossings) on~$\partial M$, whose vertex set is~$X$ and whose edges are the subpaths of~$c$ between consecutive points of~$X$ (``consecutive'' when walking along~$c$).  This is a four-valent graph, possibly with loops and multiple edges, drawn on~$\partial M$ in generic position; some edges cross (or self-cross) unless $X=C$.  A special case is when $X$ is empty; in that case, $G_X$ is a closed curve (equal to~$c$).

Alternatively, one may find useful to think of $G_X$ as follows: In the neighborhood of each crossing point~$u$ of~$c$ in $C\setminus X$, pull one of the two pieces of~$c$ passing through~$u$ slightly in the interior of~$M$, thus removing the crossing~$u$.  The result is an embedding of the graph~$G_X$, and the drawing that we consider is obtained by pushing back the pieces to~$\partial M$.

We need to carefully distinguish between objects in~$G_X$ and their images in~$\partial M$.  For this purpose, we employ the standard graph theory terminology of \emphdef{closed walk} in (the abstract graph) $G_X$.  A \emphdef{cycle} in (the abstract graph) $G_X$ is a closed walk without repeated vertices.  The image, in~$\partial M$, of a cycle in~$G_X$ may, of course, self-cross (at points in~$C\setminus X$).

Let $H$ be an abstract graph, and let $T$ an arbitrary spanning tree of~$H$.  An \emphdef{elementary cycle} in~$H$ (with respect to~$T$) is a cycle in~$H$ with all its edges in~$T$, except precisely one.  We will use the following easy and well known fact:
\begin{Fact}\label{F:elem}
Let $H$ be an abstract graph, viewed as a topological space.  For any spanning tree~$T$ of~$H$, the homotopy class of every closed walk in~$H$ lies in the normal subgroup generated by the homotopy classes of its elementary cycles.
\end{Fact}

\begin{Lemma}\label{L:elem}
  Let us assume the hypotheses of Theorem~\ref{T:main}.  Let $C$ be the set of self-crossing points of~$c$.  Then:
  \begin{enumerate}
      \item If $c$ is contractible in~$M$, then for some $X\subseteq C$, the image in~$\partial M$ of every cycle in~$G_X$ is virtually strongly contractible.
      \item If for some $X\subseteq C$ and some spanning tree~$T$ of~$G_X$, the image, in~$M$, of every elementary cycle in~$G_X$ (with respect to~$T$) is contractible, then $c$ is contractible in~$M$.
  \end{enumerate}
\end{Lemma}
\begin{proof}
  1.  Assume that $c$ is contractible in~$M$.  By Theorem~\ref{theorem:essentialloop}, there is a tower for~$c$ such that $\partial V_0$ is a disjoint union of spheres.  Moreover, $c$ lifts to a closed curve~$\tilde c$ on~$\partial V_0$.  Let $X$ be the set of self-crossing points of~$c$ that are the projections of the self-crossing points of~$\tilde c$.  The image of every cycle in~$G_X$ is virtually strongly contractible, because the image of every such cycle lifts to a simple closed curve in~$\partial V_0$, which is a sphere.
  
  2.  By Fact~\ref{F:elem}, if for some $X\subseteq C$, the image of every elementary cycle in $G_X$ is contractible in~$M$, then the image of every closed walk in~$G_X$ is contractible in~$M$.  As $c$ is the image of a closed walk in~$G_X$, this curve~$c$ is contractible in~$M$ as well.
\end{proof}

\begin{proof}[Proof of Theorem~\ref{T:main}]
  Here is the algorithm.  For every possible subset~$X$ of self-crossing points of~$c$, we consider the set~$\Gamma_X$ of the elementary cycles in~$G_X$ (with respect to an arbitrary but fixed spanning tree of~$G_X$), and apply Proposition~\ref{P:special} to each element of~$\Gamma_X$.  If, for some~$X$, the algorithm of Proposition~\ref{P:special} returns~(i) for every elementary cycle in~$G_X$ (implying that these cycles are contractible in~$M$), then we report that $c$ is contractible in~$M$; moreover, if at least one instance of the algorithm of Proposition~\ref{P:special} has returned a curve, we also return one of these output curves.  Otherwise, we report that $c$ is non-contractible in~$M$.
  
  The proof of correctness follows easily from the previous considerations.  First, if $c$ is contractible in~$M$, then Lemma~\ref{L:elem}(1) implies that for some choice of~$X$, the cycles in~$\Gamma_X$ are virtually strongly contractible; thus the algorithm in Proposition~\ref{P:special} returns~(i) for these cycles, and the above algorithm reports that $c$ is contractible in~$M$.  If the algorithm returns a curve, then it clearly satisfies the specification.  On the other hand, if the algorithm returns no curve, this means that all the elementary cycles in~$G_X$ are contractible in~$\partial M$, which implies, by Fact~\ref{F:elem}, that $c$ is also contractible in~$\partial M$, so it is correct to output no curve.
  
  On the other hand, if the algorithm reports that $c$ is contractible in~$M$, this is because, for some choice of~$X$, every elementary cycle in~$G_X$ is contractible in~$M$.  Lemma~\ref{L:elem}(2) implies that $c$ is indeed contractible in~$M$.

  Recall that the algorithm in Proposition~\ref{P:special} runs in polynomial space and $2^{O((m+n+t)^2)}$ time.  The present algorithm runs $2^m$ times the algorithm in Proposition~\ref{P:special} (these calls being independent).  Thus, it runs in polynomial space and $2^{O((m+n+t)^2)}$ time.
  \end{proof}

\section{Extensions and Refined Analysis in Special Cases}\label{section:extensions}

In this section, we show that essentially the same algorithm as in Theorem~\ref{T:main} can output more information if desired, and show that the \BdContract{} problem, when restricted to two special cases, lies in the complexity class~NP.

\subsection{Extensions}

\begin{Theorem}\label{T:mainformula}
  Consider an input to the \BdContract{} problem, and assume that $c$ is contractible in~$M$.  Let $G$ be the graph that is the image of~$c$.  In $2^{O((m+n+t)^2)}$ time, we can compute:
      \begin{itemize}
      \item $N=3\cdot2^{m/2}$ closed curves $c_1,\ldots,c_N$ in~$G$, such that each has complexity $O(m+n)$, is contractible in~$M$, and can be slightly perturbed on~$\partial M$ by smoothing some self-crossings of~$c$ to be made simple;
      \item $N'=2^{m/2}$ paths $p_1,\ldots,p_{N'}$ in~$G$, each of complexity $O(m+n)$,
       \item and a formula expressing the homotopy class of~$c$ in~$G$ as being in the normal product of the homotopy classes of $c'_1,\ldots,c'_N$ in~$G$ where $c'_i$ is the same as $c_i$ but connected to the basepoint; more precisely, it indicates that $c$ is homotopic, in~$G$, to a curve that is the concatenation of the paths $c_i$, $p_i$, and~$p_i\inv$ in some order, each appearing exactly once, such that the pairs $p_i$ and~$p_i\inv$ are conjugator paths.
      \end{itemize}
\end{Theorem}
\begin{proof}
  The proof is essentially a refinement of the proof of Theorem~\ref{T:main}.  One can refine Proposition~\ref{P:special} so that, if the answer is~(i), some paths $p_i$ and some curves~$c_i$ as indicated in the statement of the theorem are computed so that $[c]_G$ is expressed as above.  Expressing~$[c]_G$ in terms of the homotopy classes of these curves can be done explicitly, as in the proof of Lemma \ref{lemma:doublearc-noninter} or~\ref{lemma:doublearc-inter} (see also Hempel~\cite[Lemma 4.7]{Hem04}).  The formula is obtained by recurrence on the formula indicated in either Lemma~\ref{lemma:doublearc-noninter} or Lemma~\ref{lemma:doublearc-inter}.  Finally, since $c$ is contractible in~$M$, the algorithm of Theorem~\ref{T:main} returns this extended output of Proposition~\ref{P:special}, as desired.  This does not change the asymptotic time complexity of the algorithm (although the space complexity becomes exponential).
\end{proof}

\begin{Theorem}\label{T:maindisk}
  Consider an input to the \BdContract{} problem, and assume that $c$ is contractible in~$M$.  In $2^{O((m+n+t)^2)}$ time, we can compute a PL singular disk in~$M$ bounded by~$M$, of complexity $2^{O((m+n+t)^2)}$.
\end{Theorem}
\begin{proof}
  We can modify the algorithm in a way that, at each recursive call, a singular disk bounding~$c$ is computed.  Proposition~\ref{P:simple} can be modified to return an explicit PL embedded disk, by presenting the normal disk found explicitly.  Then, in Proposition~\ref{P:special}, one can also return a PL singular disk bounding the curve~$c$, if $c$ is found to be contractible.  This does not change the time complexity of the algorithm (although the space complexity becomes exponential).
\end{proof}
\subsection{Refined analysis in special cases}

It would be very interesting to know whether the contractibility problem lies in NP.  We are presently unable to decide this; the obstruction for our contractibility algorithm to be in NP is the size of the recursion tree of the algorithm of Proposition~\ref{P:special}.  Nonetheless, we show that the \BdContract{} problem is indeed in NP in two special cases.  Formally, in this section, because being in NP makes sense only for decision problems, we only consider the decision problem associated to \BdContract. 
\subsubsection{Curves with few self-intersections} 

The first special case is when the number of self-crossings is small:
\begin{Theorem}\label{T:NP}
  The decision problem associated to \BdContract{}, restricted to inputs with $m=O(\log(n+t))$, lies in the complexity class~NP.
\end{Theorem}
\begin{proof}
  Assume that $c$ is contractible.  The certificate that it is the case is the formula of Theorem~\ref{T:mainformula}, together with, for each curve~$c_i$, a normal disk bounded by (a slightly perturbed simple copy of)~$c_i$.  Because of the upper bound on~$m$, this formula has polynomial size; moreover, by the proof of Proposition~\ref{P:simple}, for each curve~$c_i$ there is such a normal disk with polynomial size.  Moreover, since the formula expresses an equality in the homotopy group of~$G$, which is a free group, it is trivial to check its validity in polynomial time in its size.  Finally, as in the proof of Proposition~\ref{P:simple} (see Agol et al.~\cite{AHT06}), in polynomial time one can check whether indeed normal disks bounding slightly perturbed simple copies of the $c_i$s are provided.
  
  (Alternatively, one could provide the set~$X$ of self-crossings in the proof of Theorem~\ref{T:main}, the recursion tree of Proposition~\ref{P:special} together with, at each internal node, the vertices $u$ and~$v$ where the curve is smoothed, and, at each leaf, a normal disk.)
\end{proof}

\subsubsection{The case of the torus boundary}

The second case in which we prove that the decision problem associated to \BdContract{} is in NP is perhaps the simplest non-trivial case, namely, the case where the boundary component of $M$ containing the curve is a torus.  Using the fact that the fundamental group of the torus is Abelian, we prove:

\begin{Theorem}\label{theorem:torus}
  The decision problem associated to \BdContract{}, restricted to the case where the connected component of~$\partial M$ containing~$c$ is a torus, lies in the complexity class~NP.
\end{Theorem}
\begin{proof}
  As above, let $G$ be the four-valent graph whose image is~$c$.  By a \emph{subcurve} of~$c$ we mean any closed curve~$c'$ whose image lies in the image~$G$, and that uses every edge of~$G$ at most once.  We first claim the following: The number of distinct homotopy classes of subcurves of~$c$ is $O((m+n)^2)$.  To see this, let $T$ be a spanning tree of~$H$, and let $G':=G/T$ be obtained from~$G$ by contracting~$T$.  Each subcurve of~$c$ becomes a walk without repeated edges in~$G'$, which is a one-vertex graph embedded on the torus.  So the loops of~$G'$ fall into at most three distinct homotopy classes, of the form $a$, $b$, and $ab$ (in multiplicative notation).  Moreover, the number of edges in~$G'$ is $O(m+n)$.  Since the fundamental group of the torus is Abelian, each subcurve of~$c$ is homotopic to a homotopy class of the form $a^pb^q$, where $|p|$ and~$|q|$ are at most $2(m+n)$.  The claim follows.

  Next, we show that there exists a certificate of polynomial length for contractibility of the curve~$c$.  We use the formula provided by Theorem~\ref{T:mainformula}.  The closed curves appearing in that formula are subcurves of~$c$; up to replacing some curves with others within the same homotopy class in~$\partial M$, and considering that the formula expresses an equality in homotopy not in~$G$ but in~$\partial M$, we can assume that there are $N=O((m+n)^2)$ curves.  Since the fundamental group of the torus is Abelian, we can omit the conjugator paths, and thus the formula simplifies as the product of the homotopy classes of $N$~closed curves, each elevated to some power that is at most $2^{m/2}$.  The first part of the certificate is this formula; since there are polynomially many curves, and since the power can be encoded in binary, this takes polynomial space.  The second part of the certificate is a normal disk spanning each of the $N$~curves; this again takes polynomial space.

  Now assume that such a certificate is given; we need to check it in polynomial time.  We first check that the formula is valid in homotopy on~$\partial M$; since the boundary component is a torus, this is easy to do in polynomial time using homology; alternatively, we can use previous results~\cite{lr-hts-12,ew-tcsr-13}.  Then, one need to check the second part of the certificate, namely, that each normal surface provided is indeed a normal disk bounding the corresponding curve; this can be done in polynomial time, as in the proof of Proposition~\ref{P:simple} (see also Agol et al.~\cite{AHT06}).
\end{proof}

\paragraph{Acknowledgments.}
The authors would like to thank Jonathan Spreer for a stimulating discussion at the early stage of this research.

\bibliographystyle{plain}
\bibliography{bib}


\end{document}